\documentclass{elsart5p}

\usepackage{graphics}
\usepackage{graphicx}
\usepackage{amssymb}
\begin{document}

\begin{frontmatter}

\title{Kondo effect in transport through Aharonov-Bohm and Aharonov-Casher
interferometers}

\author[AA]{A. M. Lobos},
\author[BB]{A. A. Aligia \corauthref{PPP}}, \ead{aligia@cab.cnea.gov.ar}

\address[AA]{DPMC-MaNEP, University of Geneva, 24 Quai Ernest Ansermet, 
CH-1211 Geneva 4, Switzerland}
\address[BB]{Centro At\'{o}mico Bariloche and Instituto Balseiro, Comisi\'{o}n Nacional
de Energ\'{\i }a At\'{o}mica, 8400 Bariloche, Argentina}

\corauth[PPP]{Corresponding author. Fax: +52 2944 445299}

\begin{abstract}
We derive the extension of the Hubbard model to include Rashba spin-orbit coupling 
that correctly describes Aharonov-Bohm and Aharonov-Casher
phases in a ring under applied magnetic and electric fields. 
When the ring
is connected to conducting leads, we develop a formalism that is able to
describe both, Kondo and interference effects. We find that in
the Kondo regime, the spin-orbit coupling reduces strongly the conductance from the unitary
limit. This effect in combination with the magnetic flux, can be used to
produce spin polarized carriers.
\end{abstract}


\end{frontmatter}

\section{Introduction}

There is a great interest in nanoscopic systems, either
because of its potential application in nanodevices or as ideal systems to
test theories for non-trivial physical problems. In particular, the Kondo
effect is present in many of these systems, like magnetic impurities on
clean noble metal surfaces or quantum corrals \cite{si}, small clusters on
these surfaces \cite{jan,tri} and systems of 
quantum dots \cite{wiel,zaf,oreg,zit,lobos06,dias}. In particular, the unitary limit,
which consists in the maximum possible conductance through a quantum dot has
been reached experimentally \cite{wiel}. The Kondo model out of equilibrium
brings new challenges to the theory \cite{grobis,none}.

On the other hand, effects of interference in quantum paths and the
Aharonov-Bohm effect have been demonstrated in mesoscopic rings with
embedded quantum dots \cite{wiel,zaf,ji,hol}.

The calculation of transport through a mesoscopic ring in which both
interference effects and interactions leading to the Kondo effect are
present is not trivial. An example of this is a recent debate about the role
of interactions in dephasing \cite{jiang}. Even knowing the exact
eigenstates of the ring, there is no simple procedure to calculate the
conductance. When the coupling $V$ of the ring to the conducting leads is
small, using perturbation theory up to second order in $V$, 
an expression has been derived, which is also exact for any $V$ in the 
non-interacting limit, the Jagla-Balseiro formula \cite {jag}. 
Similar equations were used recently \cite{hal,ihm}. Alternative
perturbative expressions were also proposed \cite{ple}. Unfortunately these
expressions are not valid in the Kondo regime, in which the ground state of
the isolated ring is Kramers degenerate (odd number of electrons), because
the ground state of the whole system is a singlet with a characteristic
energy scale (the Kondo temperature) $T_{K}\sim W\exp [-1/ \rho_0 J]$, where $W$ is
the band width, $\rho_0$ the density of states at the Fermi level
and $J \approx V^{2}$ [see Eq. (\ref{tk}) or Ref. \cite{tri}], 
which cannot be
recovered by perturbation theory in $V$. Previous
calculations of the conductance through strongly correlated rings in which
the effects of interference were important to detect signatures of
spin-charge separation, assumed that a Zeeman term destroys the Kondo effect
in the system \cite{hal}. For a ring described by the ionic Hubbard
model, it has been shown that the conductance through the system is related
to the quasiparticle weight \cite{ihm}. The physics of the Kondo
regime was recovered by mapping the model into an impurity Anderson model,
but at the cost of losing interference effects.

One of the branches of nanophysics with particular recent interest is
spintronics, which consists in developing means of creating and controlling
spin polarized currents in nanoscale systems \cite{zut}. An important
ingredient for this purposes is the Rashba spin-orbit coupling 
\begin{equation}
H_{\rm{SOC}}=\alpha \vec{\sigma}\cdot \vec{\mathrm{E}}\times (\vec{\mathrm{p}}-e\vec{\mathrm{A}}),
\label{soc}
\end{equation}
which is present in quantum wells. Recent experiments
in semiconductor mesoscopic rings have shown that the conductance oscillates
not only as a function of the applied magnetic field (Aharonov-Bohm effect)
but also as a function of the applied electric field $\vec{\mathrm{E}}$
perpendicular to the plane of the ring (Aharonov-Casher effect) \cite{kon,
ber}. In this effect, the electrons, as they move, capture a phase that
depends on the spin, as a consequence of spin-orbit coupling. The main features of the
experiment can be understood in a one-electron picture \cite{shen,mol}.
However, this picture is inadequate in the presence of strong correlations.
In particular, the Kondo effect cannot be described.

The aim of the present work is two-fold: first, to provide a method to
calculate the conductance through a ring that is able to capture at the same time the
Kondo physics and interference effects; second, to apply the formalism to
calculate the conductance through a ring under the action of both,
magnetic and electric fields together with Rashba spin-orbit coupling $H_{\rm{SOC}}$. In the
second task we stumbled with the difficulty of finding the form of the
Hubbard model in the presence of $H_{\rm{SOC}}$, which adequately
describes the periodicity with the applied electric field seen in the
experiments. We solved this problem using a non-abelian gauge transformation
in the continuum version of the model, which translates the effect of $H_{\rm{SOC}}$ 
to a change in the boundary conditions.

The main physical result is that the combination of both Kondo
effect and spin-orbit coupling leads to a strongly spin dependent conductivity that might
be used for spin filtering purposes. Some results have been recently
published \cite{abc}. In section 2 we briefly explain the general method
to calculate the conductance. In section 3, the Hubbard model including the effect
of the spin-orbit coupling Eq. (\ref{soc}) is derived. Section 4 contains 
results for a four-site ring. Section 5 is a summary and discussion. 

\section{Formalism to calculate the conductance}

To calculate the conductance through an interacting subsystem, it is
necessary to obtain some Green functions of the whole system, which includes 
the interacting part both leads \cite{meir}. 
An example of an interferometer
assembled experimentally with one quantum dot is given in Fig. 1 of Ref. 5. 
The Hamiltonian can be written as

\begin{eqnarray}
H = H_I+H_l+H_V \nonumber \\
H_I=H_{U}^{\prime }-V_{g}\sum_{i\sigma }d_{i\sigma }^{\dagger }d_{i\sigma },
\nonumber \\
H_l=t_{c}(\sum_{i=0,\sigma }^{-\infty }c_{i-1,\sigma }^{\dagger }c_{i\sigma
}+\sum_{i=1,\sigma }^{\infty }c_{i+1,\sigma }^{\dagger }c_{i\sigma }+\mathrm{%
H.c}),  \nonumber \\
H_V=V(\sum_{\sigma }c_{0\sigma }^{\dagger }d_{0\sigma }+c_{1\sigma }^{\dagger
}d_{1\sigma }+\mathrm{H.c.})\mathrm{.}  \label{h}
\end{eqnarray}
$H_I$ describes the interacting subsystem (in the case described
below, a Hubbard ring with spin-orbit coupling). The operator $d_{i\sigma }^{\dagger }$
creates an electron with spin $\sigma $ at site $i$ in this subsystem. The
second term of $H_I$ represents the effect of a gate voltage. 
The term $H_l$ describes both leads, and $H_V$ is the
coupling between both leads and the interacting region. 
We label the sites
in this region so that site 0 is coupled to the left lead and site 1 to the
right one. The prime in $H_{U}^{\prime }$ reminds us that the Hamiltonian of
the subsystem is written in a particular gauge, as described in the next
section. Note that the phases of all $c_{i\sigma }$ operators can be chosen
so that $t_{c}$ and $V$ are real, regardless of the above mentioned gauge.
The Zeeman term is neglected here.
To reduce the number of parameters, we have assumed identical left and right 
leads and the same coupling between any of them and the interacting subsystem.

Since $H_I$ describes a finite system, it can be diagonalized
exactly. Our general approach to calculate the conductance amounts to a
truncation of the Hilbert space of $H_I$, retaining only two
neighboring charge configurations, with $n$ and $n-1$ particles. It can be
shown that this procedure is valid for small enough $V$ \cite{lobos06}.
Calculating the matrix elements of $H_{V}$ in the truncated Hilbert space
leads to a generalized Anderson model 
\begin{eqnarray}
H_{\rm{GAM}}=H_{l}+\sum_{n,j}E_{j\sigma }^{n}|\psi _{j}^{n}\rangle \langle
\psi _{j}^{n}|
\nonumber \\
+V\sum_{\eta k,j\sigma }(\beta _{kj\sigma }^{\eta }c_{\eta
\sigma }^{\dagger }|\psi _{k}^{n-1}\rangle \langle \psi _{j}^{n}|+\mathrm{%
H.c.)},  \label{gam}
\end{eqnarray}
where $|\psi _{j}^{n}\rangle $ and $E_{j}^{n}$ denote the $j$-th eigenvector
and eigenvalue of $H_I$ in the configuration with $n$ particles
and 
\begin{equation}
\beta _{kj\sigma }^{\eta }=\langle \psi _{k}^{n-1}|d_{\eta \sigma }|\psi
_{j}^{n}\rangle ~~ (\eta=0,1).  \label{beta}
\end{equation}

In the simplest case, only the spin singlet (doublet) ground state of $H_I$ is
relevant for the configuration with an even (odd) number of electrons and $%
H_{\rm{GAM}}$ reduces to the ordinary Anderson model \cite{ihm}. However,
interference effects, for example the vanishing of the conductance through
the ring for certain values of the flux \cite{hal} is a consequence of an 
\emph{orbital} degeneracy of levels for that flux and more states than one
Kramers doublet should be included in $H_{\rm{GAM}}$, as we report in section 4.

In general, $H_{\rm{GAM}}$ can be represented using one or more slave
bosons that represent the relevant eigenstates of $H_I$ with an
even number of particles \cite{cox,trip}. A model with a doublet hybridized
with a singlet and a triplet has been solved using the numerical
renormalization group \cite{allub}. From the solution of $H_{\rm{GAM}}$,
the conductance is obtained using known expressions that relate it
with the exact Green functions of $H_{\rm{GAM}}$ \cite{meir}.

\section{Hubbard model in the presence of spin-orbit interaction}

The simplest way to add the physics of the spin-orbit coupling in the Hubbard model seems to
be to replace $H_{\rm{SOC}}$ given by Eq. (\ref{soc}) by a tight-binding version 
in which the current in each link is multiplied by the Pauli matrix perpendicular
to the link and to the electric field, to build the cross product of Eq. (\ref{soc}) \cite {rey}. 
For an electric field in the $z$ direction
and a ring in the $x,y$ plane this gives:

\begin{eqnarray}
H_{tb} &=&\alpha E_{z}\hbar /(2a)\sum_{i}[i\cos \varphi (d_{i+1\uparrow
}^{\dagger }d_{i\downarrow }+d_{i+1\downarrow }^{\dagger }d_{i\uparrow }) \nonumber \\
&&+\sin \varphi (d_{i+1\uparrow }^{\dagger }d_{i\downarrow
}-d_{i+1\downarrow }^{\dagger }d_{i\uparrow })+\mathrm{H.c.}]. \label{tb}
\end{eqnarray}
However, in the experimentally assembled \emph{rings} it is clear that the
conductance \emph{oscillates} as $\vec{\mathrm{E}}$ increases \cite{kon, ber}
and therefore one expects that the electric field enters an exponential as a phase,
and this is not apparent in Eq. (\ref{tb}). 

To follow a procedure that
leads to such an exponential dependence, we consider the version of the model
in the continuum $H_{U}^{c}$. This is obtained from the
noninteracting version derived by Meijer {\it et al.} \cite
{mei} for electric and magnetic fields perpendicular to the plane of the
ring, adding a local interaction. The Hamiltonian is

\begin{eqnarray}
H_{U}^{c}=\hbar \Omega \sum_{j}\left[ -i\frac{\partial }{\partial \varphi
_{j}}-\frac{\phi }{\phi _{0}}+\frac{\gamma }{2}\sigma _{r}(\varphi
_{j})\right] ^{2}
\nonumber \\
+U\sum_{i<j}\delta (\varphi _{i}-\varphi _{j}),  \label{hu}
\end{eqnarray}
where $\varphi_j$ is the azimuth of the $j$-th electron, 
$\Omega =\hbar /(2m^{*}r^{2})$, $m^{*}$ is the effective electron
mass, $r$ is the radius of the ring, $\gamma =\alpha E_{z}/(r\Omega )$
is proportional to the Rashba constant $\alpha$ and the electric field $E_{z}$, 
$\phi =B\pi r^{2}$ is the magnetic flux, $\phi _{0}=hc/e$ is the flux quantum
and $\sigma _{r}(\varphi )=$ $\sigma _{x}\cos {\varphi }$ + $\sigma _{y}\sin 
{\varphi }$ is the Pauli matrix in the radial direction.

It can be easily checked that the unitary transformation 
$T=\prod_{j}t(\varphi _{j})$, with

\begin{eqnarray}
t(\varphi ) &=&\mathrm{exp}\left[ -i\sigma _{z}\frac{\varphi }{2}\right] 
\mathrm{exp}\left[ i\vec{\sigma}.\vec{n}_{\theta }\frac{\varphi ^{\prime }}{2%
}\right] \mathrm{exp}\left[ i\frac{\phi }{\phi _{0}}\varphi \right] \rm{, }
\nonumber \\
\vec{n}_{\theta } &=&(-\sin {\theta },0,\cos {\theta })\rm{, }\theta
=\arctan {(\gamma )}\rm{, }  \nonumber \\
\varphi ^{\prime } &=&\varphi \sqrt{1+\gamma ^{2}},  \label{t}
\end{eqnarray}
makes the dependence of the fields [$\phi $ and $\gamma$ in 
Eq. (\ref{hu})] disappear in the transformed Hamiltonian $H_{U}^{c\prime }=T^{\dagger
}H_{U}^{c}T$. Therefore $H_{U}^{c\prime }$ is the continuum version of the
ordinary Hubbard model, in which the magnetic flux and the spin-orbit coupling have been gauged
away. The price to pay is that the transformed one-particle spinors satisfies the boundary
conditions $\chi ^{\prime }(2\pi )=t^{\dagger }(2\pi )\chi ^{\prime }(0)$
(instead of periodic ones). This implies that in the Hubbard model for a
ring of $N$ sites, the last hopping (from angle $\varphi =-2\pi /N$ to 0)
should be modified. Diagonalizing $T^{\dagger }(2\pi )$ one obtains the
eigenvalues $\exp [i(\Phi _{AB}+\sigma \Phi _{AC})]$, where $\sigma =\pm 1$
for spin pointing in the $\pm \vec{n}_{\theta }$ direction, $\Phi _{AB}=2\pi
\phi /\phi _{0}$, and $\Phi _{AC}=\pi \{[1+\gamma ^{2}]^{1/2}-1\}$.

Then, the transformed Hubbard model for a ring of $N$ sites takes the form

\begin{eqnarray}
H_{U}^{\prime } &=&-\sum_{i=0,\sigma }^{N-2}t\left[ d_{i+1\sigma }^{\dagger
}d_{i\sigma }+\mathrm{H.c.}\right]   \nonumber \\
&&-t\left[ e^{i(\Phi _{AB}+\sigma \Phi _{AC})}d_{0\sigma }^{\dagger
}d_{N-1,\sigma }+\mathrm{H.c.}\right]   \nonumber \\
&&+U\sum_{i}d_{i\uparrow }^{\dagger }d_{i\uparrow }d_{i\downarrow }^{\dagger
}d_{i\downarrow },  \label{hup}
\end{eqnarray}
where

\begin{equation}
\Phi _{AC}=\sqrt{\pi ^{2}+R^{2}}-\pi ,  \label{phiac}
\end{equation}
and the ratio $R=\pi \gamma $ is given by $\hbar \alpha E_{z}N/(2ta)$
if the mass is eliminated from the curvature of the bottom of the non-interacting
band ($m^{*}=\hbar ^{2}/(2ta^{2})$, where $a=2\pi r/N$ is the lattice
parameter). If instead $m^{*}=\hbar k_{F}/v_{F}$ is used, where $k_{F}$, $v_{F}$ 
are the Fermi wave vector and velocity respectively
\begin{equation}
R= \frac{\hbar \alpha E_{z}N k_{F}a }{2ta \sin (k_{F}a)}.  \label{r2}
\end{equation}
In any case, the same
qualitative physics is described. The rotational invariance of 
$H_{U}^{\prime }$ can be recovered replacing 
$d_{j\sigma }^{\dagger }=\tilde{d}_{j\sigma }^{\dagger }\exp 
[i(\Phi _{AB}+\sigma \Phi _{AC})j/N]$.

It is important to remark that the spin quantization axis of Eq. (\ref{hup})
varies with position. For $\varphi =0$ (corresponding to $y=0$), it is
given by $\vec{n}_{\theta }$ (see Eq. (\ref{t})) and it lies in the $x,z$ plane. For
other angles the spin quantization axis rotates with $\varphi $ in such a
way that the component in the plane of the ring ($x,y)$ always points
towards the center of the ring.

It is interesting to note that Berry phases captured in adiabatic evolutions
of a system in $\Phi _{AB}$ and $\Phi _{AC}$ give information on the
polarization, opening of a spin gap, phase transitions and ferrotoroidic
moments \cite{abb,bat}.

\section{Results for 4-site ring}

For the explicit calculation, we take a ring of $N=4$ sites, connected to
the leads at opposite sites, lying at $\phi=0$ and $\phi=\pi$.
We assume that the leads are described by a
constant density of states $\rho _{0}=1/W$ and set the band width $W=60t$
(much larger than the hopping $t$ in the ring). The Fermi energy of the leads is
set at the on-site energy in the ring (zero).We have included all doublet
states with $n=3$ and the singlet ground state for $n=4$ of $H_I$, to construct
the generalized Anderson model $H_{\rm{GAM}}$
(see Eq. (\ref{gam})). We have solved $H_{\rm{GAM}}$ in a slave boson
mean-field approximation, which is known to reproduce correctly the exponential
dependence of the Kondo scale on the parameters. Details are given in Ref. \cite{abc}.

\subsection{The different regimes}

\begin{figure}[!ht]
\includegraphics[width=0.43\textwidth]{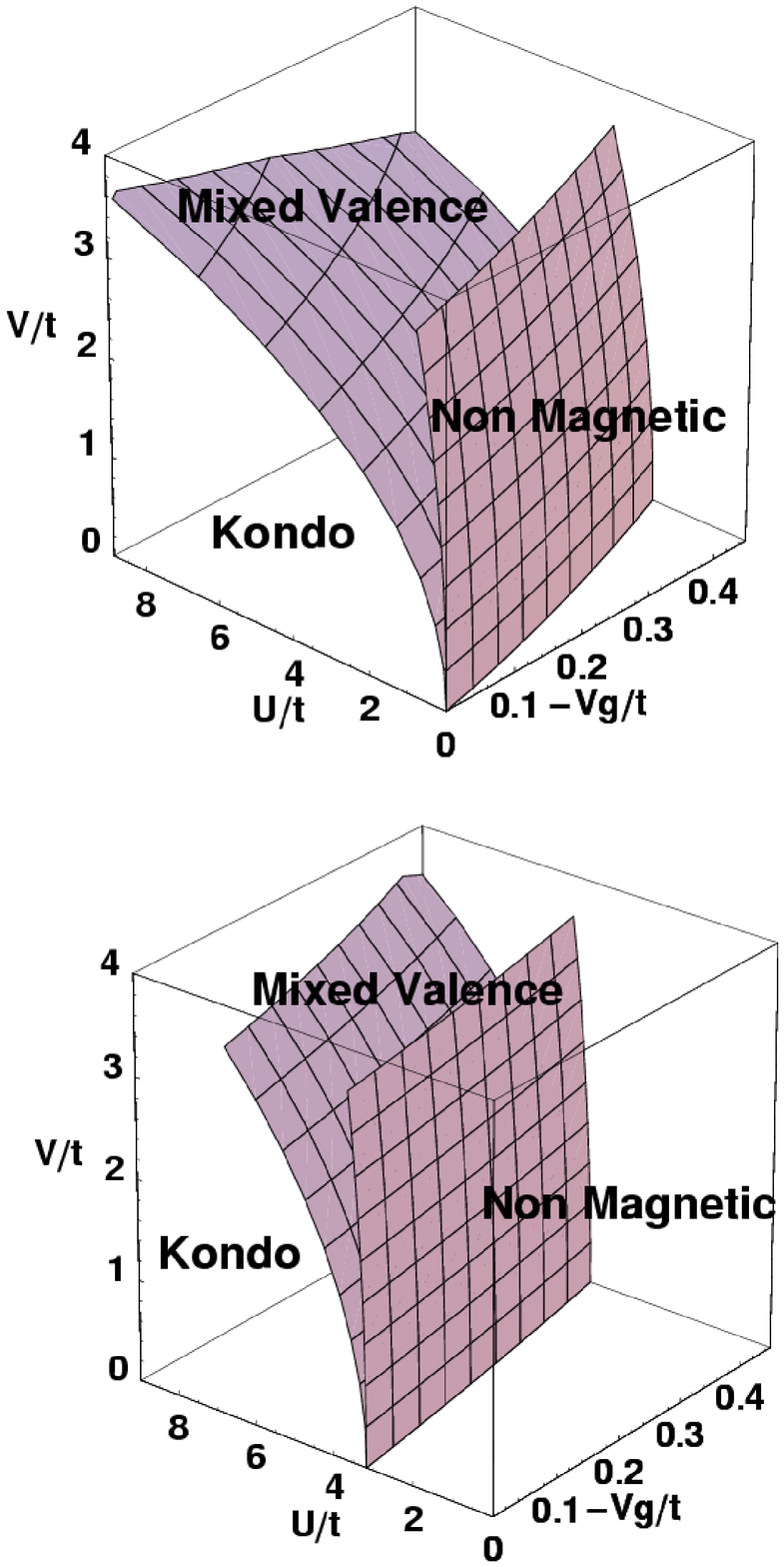}
\caption{Different regimes of the model for spin-orbit strength $R=0$ and two values
of the applied magnetic flux:$\Phi _{AB}=0$ (top) and $\Phi _{AB}=\pi $
(bottom).}
\label{bf1}
\end{figure}

The properties of the effective generalized Anderson model, Eq. (\ref{gam})
and therefore the conductance through the ring, differ according to
different regimes which depend on the ratio of the charge-transfer energy $%
E_{0}^{(4)}-E_{0}^{(3)}$, where $E_{0}^{(n)}$ is the ground state of 
$H_I$ in the subspace of $n$ electrons, and the effective
resonant level width 
\begin{equation}
\Delta =\pi \rho _{0}V^{2}\sum_{\eta}|\beta_{00\sigma }^{\eta }|^{2} .  \label{delta}
\end{equation}

If the four-particle singlet is well below the
lowest Kramers doublet ($E_{0}^{(3)}-E_{0}^{(4)}\gg \Delta $) the system is
in the non-magnetic regime. For increasing values of $E_{0}^{(4)}$, the
system enters first the intermediate valence zone 
($|E_{0}^{(3)}-E_{0}^{(4)}|\sim \Delta $), 
and then the Kondo regime when the lowest spin doublet
is well below the singlet ($E_{0}^{(4)}-E_{0}^{(3)}\gg \Delta $).

In Fig. \ref{bf1} we show the different regions of parameters of the model $U
$, $V_{g}$ and $V$ (setting $t$ as the unit of energy) corresponding to the
different regimes in absence of spin-orbit coupling. The boundary between the mixed-valence
regime and the nonmagnetic (Kondo) one has been defined arbitrarily in the
figure by 
$E_{0}^{(3)}=E_{0}^{(4)}+\Delta $ ($E_{0}^{(3)}=E_{0}^{(4)}-8\Delta $). 
Note that there is a strong dependence of the boundaries with the
applied magnetic flux. The four-particle singlet is favored for $\Phi _{AB}=\pi 
$. This can be understood already for the non-interacting ring, since the
ground-state energy for four particles $E_{0}^{(4)}$ is optimized for $\Phi
_{AB}=\pi $, while it passes through a relative maximum at $\Phi _{AB}=0$.
In the strongly interacting case $U\rightarrow +\infty $, $H_{U}^{\prime }$
reduces to a $t-J$ model and $E_{0}^{(4)}$ becomes independent of the flux,
while $E_{0}^{(3)}$ is minimized for $\Phi _{AB}=0$ \cite{hal}.

Clearly, there is no mixed valence regime for $V=0$, since this implies 
$\Delta =0$, while the extension of this regime increases nearly
quadratically with $V$.

\subsection{The non-magnetic regime}

\medskip
\begin{figure}[!ht]
\centering 
\includegraphics[width=0.42\textwidth]{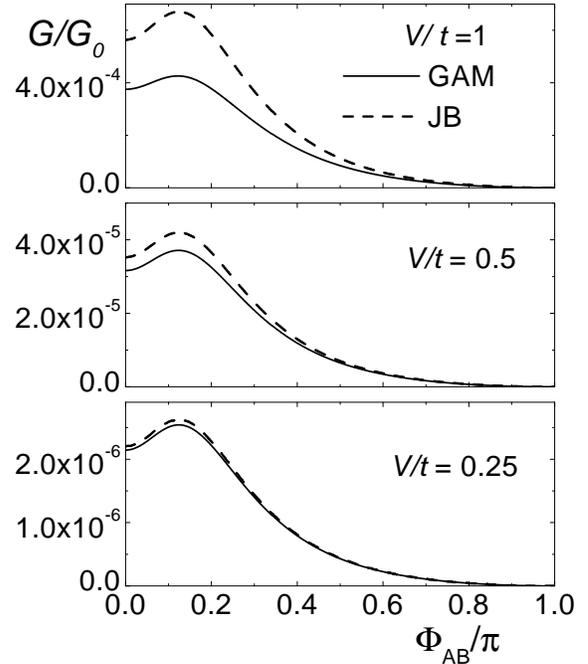}
\caption{Conductance as a function of magnetic flux for $R =0$, 
$V_{g}=-0.8t$, $U=2t$ and several values of $V$. Full lines: our formalism.
Dashed lines: Jagla-Balseiro formula.}
\label{bf2}
\end{figure}  

In Fig. \ref{bf2} we show the conductance $G=G_{\uparrow
}+G_{\downarrow }$ as a function of magnetic flux in the non-magnetic regime
for different values of $V$ and without applied electric field ($R=0$). The
quantum of conductance is denoted as $G_{0}=2e^{2}/h$ . The total
conductance is even with flux $\phi $ and we show $G$ in the
interval $0<\phi <\phi _{0}/2$ (or $0<\Phi _{AB}<\pi $). 

The conductance is very small due to the fact that there are no available
states of the ring near the Fermi energy. $G$ increases with the hopping to the leads $V$,
since this term promotes electrons to the leads at the Fermi level. In
addition, in this regime correlations play a minor role and one expects that
the Jagla-Balseiro formula \cite{jag}, which is exact in the
non-interacting case, gives accurate values for the conductance. In fact,
our results show the same qualitative behavior and for small $V$ it can be
demonstrated that both approaches are equivalent in this regime. 
However, there are significant quantitative differences for $V/t=1$.
It is difficult to state which approach is the most acurate in this case.

In absence of spin-orbit coupling, for an applied flux of half a flux quantum, the
conductance vanishes due to destructive interference, as a consequence
of reflection symmetry of the ring for a non-degenerate ground state \cite{hal}.

\subsection{Spin dependent conductance in the Kondo regime}

\begin{figure}[!ht]
\centering 
\includegraphics[width=0.43\textwidth]{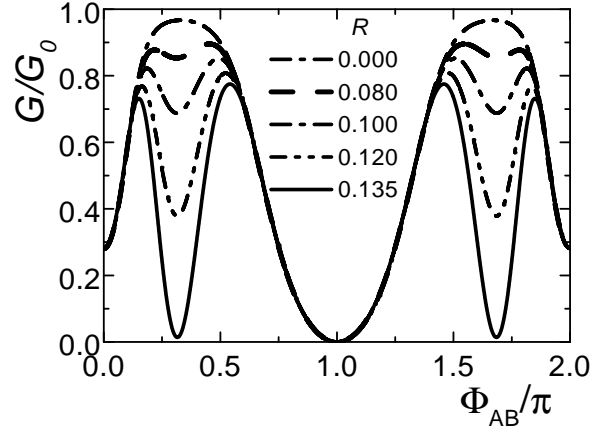}
\caption{Conductance as a function of flux for 
$V=3.5t$,$V_{g}=0$, $U=6t$  and
different values of the electric field times spin-orbit coupling strength $R$ (see text). }
\label{bf3}
\end{figure}

In the rest of this paper, we consider $E_{0\sigma }^{(3)}<E_{0}^{(4)}$,
when the ring is the mixed valence or Kondo regime. In Fig. \ref{bf3} we
show the conductance as a function of the applied magnetic flux $\Phi _{AB}$
for several values of the applied electric field $E_{z}$. We discuss first
the case $E_{z}=R=0$. In contrast to the results in the non-magnetic regime,
the conductance takes appreciable values, being near to the ideal one for $%
0.2<\Phi _{AB}<0.5$ and $1.5<\Phi _{AB}<1.8$. The system can be considered
to be in the Kondo regime for these values of $\Phi _{AB}$ and in the mixed
valence regime for other fluxes. This is also consistent with the calculated
occupation of the levels, shown elsewhere \cite{abc}, which is nearly one in
the Kondo regime. As expected from Fig. \ref{bf1}, the region of $\Phi _{AB}$
for which the system is in the Kondo regime increases with decreasing $V$,
and for $V=3.1t$ nearly perfect conductance is obtained for $0.1<\Phi
_{AB}<0.7$ and $1.3<\Phi _{AB}<1.9$ \cite{abc}. This corresponds to
the unitary limit. This limit, observed experimentally in systems with one quantum dot
different than ours \cite{wiel}, is characteristic of the Kondo regime. 
The Jagla-Balseiro formula fails to describe this case, giving very small values of the conductance.
Note that the conductance still vanishes at $\Phi _{AB}=\pi $
as a consequence of interference. This together with the ideal conductance
for the Kondo regime shows that our approach includes both, interference
and Kondo effects, in
contrast to previous calculations \cite{hal,ihm}.

\medskip
\medskip
\medskip
\medskip

Precisely in the Kondo regime, where the conductance reaches its maximum
possible value, is where the effect of the spin-orbit coupling times the applied electric
field $E_{z}$ is most noticeable. As it can be seen in Fig. \ref{bf3},
moderate values of $E_{z}$ are enough to decrease the conductance from $G_{0}
$ to very small values. The main reason for this is that $E_{z}$ affects
both spin projections in opposite ways breaking SU(2) symmetry, and therefore 
tends to destroy the Kondo
singlet. The stabilization energy of this singlet is of the order of the
Kondo temperature 
\begin{equation}
T_{K} \sim W\exp (-\pi (E_{0}^{(4)}-E_{0}^{(3)})/\Delta),  \label{tk}
\end{equation}
and therefore decreases exponentially inside the Kondo regime. For the
parameters of Fig. \ref{bf3}, the system is deeper in the Kondo regime for 
$\Phi _{AB}\sim \pm 0.3\pi $ (mod $2\pi $). Due to the loss of reflection
symmetry, in presence of spin-orbit coupling, complete conductance cancellation is not
realized at $\Phi _{AB}=\pi $. 

\begin{figure}[!ht]
\centering 
\includegraphics[width=0.45\textwidth]{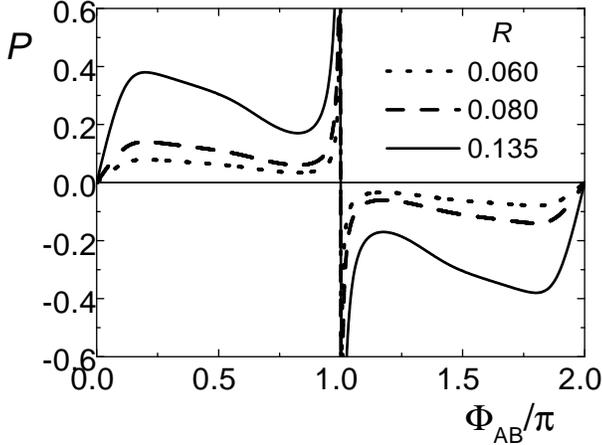}
\caption{ Polarization of the conductance 
$P=(G_{\uparrow }-G_{\downarrow })/(G_{\uparrow }+G_{\downarrow})$ 
as a function of flux for 
$V=3.5t$,$V_{g}=0$, $U=6t$ and different  
electric fields.}
\label{bf4}
\end{figure}

The above mentioned breaking of spin SU(2) symmetry as a consequence of the
spin-orbit coupling leads to significant differences in the conductance $G_{\sigma }$ for
both spin orientations $\sigma $. Since in our Hamiltonian, the quantization
axis depends on the position of the electron, this conductance should be
interpreted in the following way: if a spin $\sigma $ (up or down) in the
quantization direction $\vec{n}_{\theta }=(-\sin {\theta },0,\cos {\theta })$
(see Eq.(\ref{t})) is injected in the ring at the right lead ($\varphi =0$)
it comes out at the left lead ($\varphi =\pi $) with spin $\sigma $ in the
direction $\vec{n}_{\theta }^{\prime }=(\sin {\theta },0,\cos {\theta })$ or
vice versa. 
For an arbitrary direction, the incident wave should be
decomposed in the components along the corresponding quantization axis. 
In Fig. \ref{bf4} we show the degree of polarization of the
conductance 
$P=(G_{\uparrow }-G_{\downarrow })/(G_{\uparrow }+G_{\downarrow})$ 
as a function of $\Phi _{AB}$ for several $E_{z}$. The ratio $P=\pm 1$
at symmetric points near $\Phi _{AB}=\pi $ because one of the $G_{\sigma }$
vanishes there and the other one at the symmetric point around $\pi $ [the
Hamiltonian is invariant under time reversal and change of sign of $\Phi
_{AB}=2(\pi r)^{2}B/\phi _{0}$]. For fluxes corresponding to the Kondo
regime $P\sim 0.4$ and one of the $G_{\sigma }$ can still be near the ideal
one.

\medskip

\section{Summary and discussion}

We propose an approach to calculate the conductance through a ring of
interacting quantum dots, weakly coupled to conducting leads, that takes into account
both interference effects and non-perturbative many-body ones, mapping the
relevant states into an effective generalized Anderson model containing
several non-degenerate states. Here we have solved the model using a
slave-boson representation in the saddle-point approximation, but
alternative treatments like the numerical renormalization group are also
possible.

We have derived an extension of the Hubbard model that includes spin-orbit
interaction, absorbing it in opposite Aharonov-Casher phases for spin up and
down in an adequately chosen quantization axis, defined by a non-abelian
gauge transformation. For a non-interacting system, it has been noticed previously
that the Rashba spin-orbit coupling can be gauged away in a ring \cite{meir2}, although
the explicit dependence of the boundary conditions [the Aharonov-Casher phase 
$\Phi _{AC}$, see Eq. (\ref{phiac})]
with the applied electric field was not given. Using Bethe ansatz, the persistent
currents in a Hubbard ring were calculated as a function of $\Phi _{AC}$ 
by Fujimoto and Kawakami \cite{fuji}. These authors also note that the trick used to
gauge away the spin-orbit coupling can be extended to any SU(2) invariant
interactions (not only local as $U$). For example a nearest-neighbor repulsion 
$\sum_{i \sigma \sigma' }d_{i+1 \sigma }^{\dagger }d_{i+1 \sigma }d_{i \sigma' }^{\dagger
}d_{i \sigma'}$
is not modified by a spin rotation at any site. Therefore, the thermodynamic 
properties of the  extended Hubbard model including this interaction should not
vary with spin-orbit coupling in the thermodynamic limit, in which boundary
conditions are irrelevant. This fact is not obvious in alternative treatments \cite{gri}.

We have calculated the conductance through a ring described by the 
Hubbard model with Rashba spin-orbit coupling
in presence of magnetic and electric fields. The
main effect of the spin-orbit coupling is to tend to destroy the Kondo
effect, leading to a strong spin dependence of the conductance.

\section*{Acknowledgments}

One of us (AAA) is indebted to Liliana Arrachea, Karen Hallberg 
and Bruce Normand for useful discussions.
This investigation was sponsored by PIP 5254 of CONICET and PICT 2006/483 of
the ANPCyT. A.A.A. is partially supported by CONICET.

\end{document}